%% file: Physrev.tex
\documentclass[aps,prd]{revtex4} %Physrev 12.16.03. APS prefers one-column.
\usepackage[dvips]{graphicx} %needs [dvips] to standardize graphics inclusion (p382).
\usepackage{amssymb} %additional AMS symbols added to revtex4
\newcommand{\be}{\begin{equation}} 
\newcommand{\ee}{\end{equation}}
\newcommand{\bea}{\begin{eqnarray}}    %* would omit eqn numbers.
\newcommand{\eea}{\end{eqnarray}}
\newcommand{\bitem}{\begin{itemize}}
\newcommand{\eitem}{\end{itemize}}

\newcommand{\benum}{\begin{enumerate}} 
\newcommand{\eenum}{\end{enumerate}}
\newcommand{\bc}{\begin{center}}
\newcommand{\ec}{\end{center}}
\newcommand{\omq}{\Omega_Q}
\newcommand{\ombe}{\Omega_B}

\newcommand{\dub}{w_Q}

\newcommand{\rhoq}{\rho_Q}
\newcommand{\rhob}{\rho_B}

\begin{document}
\title{\bf WHAT WE ALREADY KNOW ABOUT QUINTESSENCE}
\author{Sidney Bludman}\email[]{bludman@mail.desy.de}
\affiliation{Deutsches Elektronen-Synchrotron
DESY,Hamburg\\University of Pennsylvania, Philadelphia}
\date{\today}

\begin{abstract}                     

Good tracking requires significant quintessence energy fraction, even in the past, 
but a potential energy that is not
yet truly slow-rolling.
The supernova bound on cosmic acceleration
excludes constant equation of state and inverse power potentials, but allows the
SUGRA potential and other
good trackers, in which quintessence energy domination and kinetic
energy suppression {\em both} began only recently.
This makes the time in which we live special in {\em two} respects.  

\end{abstract}

\maketitle   %Formats the front material.%%%\be..\ee make display eqns;
%blank lines will display eqns in .tex, but will not compile in latex.
\section{THE DARK ENERGY DENSITY IS NOW EXACTLY OR NEARLY STATIC}
\subsection{Kinematics of the Expanding and Accelerating Universe}
Supernovae Ia, cosmic shear, and angular diameter measurements all
directly explore the space-time geometry by measuring the
luminosity distance $d_L(z)=(1+z)\eta$ or the angular-diameter
distance $d_A(z)=\eta/(1+z)$, from which the comoving distance
\be\eta\equiv c\int_0^z dz'/H(z')=c\int_0^t dt'/a(t')\ee to
individual distant supernovae, chosen to be standard candles, or
to distant galaxies is inferred. (The conformal coordinate
distance to the horizon, $\eta$, describes the proper time
evolution of the scale factor $a$.) Assuming a homogeneous and
isotropic (Robertson-Walker) flat universe, the Friedmann
expansion rate is
 \be 8\pi G\rho=3 H^2 , \qquad \mbox{where} \quad H \equiv \dot
a/a. \ee  Quantum field theory requires that the energy density,
and therefore $G$, be positive, so that we can write $\sqrt{8\pi
G}\equiv\varkappa\equiv 1/M_P$, where $M_P=2.44e18~GeV$ is the
reduced Planck mass. The derived quantity, the cosmological fluid
pressure $P=-d (\rho c^2 a^3)/d a^3 $, may be positive or
negative.

In terms of geometrical quantities, \be\varkappa^2 P/c^2=-(2\dot
H+3 H^2) ,\ee the enthalpy is \be \varkappa^2(\rho+P/c^2)=-2\dot
H=-d H^2/d N=-(\varkappa^2 /3)(d\rho/dN), \ee and the over-all
barotropic index is \be\gamma \equiv -d\ln\rho/3
dN=(\rho+P/c^2)/\rho= -\frac{2}{3}(d\ln H/dN).\ee Here the
logarithm of the cosmological scale factor $N\equiv \ln
a=-\ln(1+z)$, so that $dN=H dt$. This equation of state and its
quintessence component do not depend on the Hubble radius
$H^{-1}$, but on its derivative i.e. not on the comoving distance
$\eta$, but on $d\eta/dz, ~d^2\eta/dz^2$. In cosmologies
satisfying the Weak Energy Condition $\rho+P/c^2>0$, $\dot
H<0$,
so that the expansion is monotonic. We will not
consider alternate gravity theories, such as extra dimensions, in
which the Friedmann equation (2) is modified.

In cosmology, the Ricci scalar is \be R\equiv -6(\dot H+2
H^2)=- \varkappa^2 (\rho-3 P),\ee while
the acceleration, $\ddot a/a=-\varkappa^2 (\rho+3 P)/6$, so that
\be \ddot a a/{\dot a}^2=1- d(H^{-1})/d t=-(1+3
w)/2 \ee ranges from $-2$ to $1$, when the overall equation of
state $w\equiv P/\rho$ ranges from $1$ to $-1$ . We now know that, until about red-shift $z\sim
0.7$, attractive gravity dominated the cosmological fluid so that
large scale structures could form. Only recently, after the
expansion rate in equation (7) outpaced the growth of the Hubble
radius $H^{-1}$, did $P/\rho=\dub\omq<-1/3$, and the expansion
become accelerated.  Interestingly, this direct observation of
acceleration in the present universe supports the
possibility of an inflationary early universe, which is otherwise
not directly observable.

Because the barotropic index (5) and its quintessence component
(13) depend on the first and second derivatives of the comoving
distance $\eta$, the quintessence evolution $\dub(z)$ depends on
first and second derivatives of the observed luminosity distances
\cite{Moar}. In practice, quintessence is appreciable only for
small red-shift. This means that, before $\dub(z)$ can be
determined, the inherently noisy luminosity distance $d_L(z)$ data
must be parametrized. For this, and other reasons, along with a
large number of high red-shift supernovae, precise knowledge of
other cosmological parameters will be needed
\cite{Naka,Hut,Frieman}, and can still determine only one or two
parameters characterizing the potential, such as
$w_{Q0},~(d\dub/dz)_0$.

While programs to measure luminosity and angular diameter
distances are underway, we already know that we live at a time
when \be\Omega_{Q0}=0.71\pm 0.07 , \qquad
\widetilde{\dub}<-0.78~(95\%~CL), \qquad h\equiv H_0/100=0.72\pm
0.05 ,\ee that the radiation/matter equality took place at
red-shift $z_{eq}=3454^{+385}_{-392}$
\cite{Spergel,Tonry,Bean,Coras,Efst,Bond,Teg,Wang}, dark energy
began dominating over matter $\gtrsim 6.3~Gyr$ ago, and the
cosmological expansion has been accelerating since red-shift
$z\sim 0.7$ \cite{Blud,Tonry}. The background density, \be
\rho_B=(11.67 a+0.003378)/a^4\qquad meV^4 ,\ee is now
$\rho_{B0}=11.67 ~meV^4$ and was $\rho_{Bi}=0.003378~GeV^4$ at
fiducial red-shift $z=10^{12}$.  The supernova observations fit an
average
\[\widetilde{\dub}(N)\equiv\int_0^N \dub(N')\,d N'/N ,\]
over a small range in $z$, in which the quintessence field and
$\dub (z)$ change relatively little, so that an approximate bound
on $w_{Q0}$ is the fitted $\widetilde{\dub}<-0.78$. The CBR
anisotropy and mass fluctuation spectrum, on the other hand,
depend on "early quintessence" \cite{Cald}, back to the last
scattering surface $z\sim 1100$. Where needed, we will fix
$h^2=1/2$, so that the present critical density and smooth energy
density are $\rho_{cr0}=40.5~meV^4,~\rho_{Q0}=28.8~meV^4$.

We will show that these observational constraints (8) allow only crawling quintessence
(Section III) or potentials with large current curvature (Section
IV), ``cross-over quintessence'' \cite{Wett2}. We go beyond the
many earlier optimistic treatments of the Attractor Condition
\cite{CDS,SWZ,ZWS,Fer,Liddle,Cope,Ng} and of inverse power
potentials \cite{Ratra,Wett}, to consider (1) poor trackers, (2)
post-tracker behavior in the present quintessence-dominated era,
(3) the range of initial conditions that would lead to tracking,
and (4) the numerical problems encountered in cosmological
dynamics, particularly in the freezing and tracking epochs. But
first we will review (Section II) the Attractor Condition, in
order to show how the basin of attraction shrinks for potentials
satisfying the observational constraints (8).
\subsection{Quintessence Dynamics: Potentials Not Yet Truly Slow-Rolling}

The universe is flat, presently dominated by smooth energy, and
recently accelerated. Canonical quintessence models the smooth
energy  dynamically by a spatially homogeneous light classical
scalar field, with canonical kinetic energy $K=\dot\phi^2/2$,
minimal gravitational coupling, zero true cosmological constant,
rolling down its self-potential $V(\phi)$.  This quintessence
field generically has good-or-bad attractor properties \cite{Ratra,Wett}, 
making the present universe more-or-less 
insensitive to a broad range initial conditions. Tracking quintessence was invoked to use this
attractor property to explain the small cosmological constant or
present smooth energy density, without fine tuning of the
potential or initial conditions. 
In canonical quintessence, the scalar field equation of motion \be
\ddot\phi+3H\dot\phi+dV/d\phi=0,\ee has the first integral \be
V(N)=\rho_Q+d\rho_Q/6 dN=\rho_Q(1-\dub)/2, \ee where the energy
density and pressure are
 \be \rhoq=\dot\phi^2 /2+V(\phi), \qquad P/c^2=\dot\phi^2
 /2-V(\phi),\ee
and the quintessence barotropic index  \be \gamma_Q(N)\equiv -d\ln
\rhoq/3 dN=((\rho+P/c^2)/\rho)_Q \equiv 1+\dub \ee lies between 0
and 2. Because the scalar field does not cluster on supercluster
scales, its mass must be $\lesssim 10^{-31}~eV$, which is
incredibly small.

From the energy integral (12), $\dot{\phi}^2=\gamma_Q\rhoq$ or $
(\varkappa d\phi/dN)^2=6 x^2$, where $x^2\equiv
\dot\phi^2/2\rho=\gamma_Q \Omega_Q/2$ is the quintessence kinetic
energy fraction of the total energy density. The ratio of
kinetic/potential energy $K/V=(1+\dub)/(1-\dub)$ has the rate of
change $ d\ln(K/V) dN=6(\Delta-1)$, where $\Delta(N)\equiv -d\ln
V/3\gamma_Q dN$. Thus, the {\em roll}, \be\lambda\equiv -d\ln
V/\varkappa d\phi= \sqrt{3\gamma_Q/\Omega_Q}\cdot \Delta, \ee and
\be d\dub/dN =3(1-\dub^2)(\Delta-1), \qquad \varkappa d\phi/d
N=\sqrt{3\gamma_Q\Omega_Q}\ee is a two-element non-autonomous
system for the dependent variables $\phi,~\dub$. Integrating the
second equation (15) implicitly relates $\phi$ and $V(\phi)$, so
that, if the equation of state $\dub (N)$ can be observed, the
potential can be reconstructed. The overall equation of state is
\be w=\gamma-1=w_B\ombe+\dub\omq,\ee where the dimensionless
ratios, $\omq, ~\ombe\equiv \rhob/(\rho+\rhoq)$, are the energy
density fractions in quintessence and in the background, and
$\gamma_Q,~\gamma_B \equiv -d \ln \rhob /3 dN $ are their
corresponding barotropic indices. Defining the potential energy
fraction $y(N)^2 \equiv V_Q/\rho$, equations (15) have scaling
solutions, when $\gamma_Q\approx const$: kination, when
$\gamma_Q=2,x>>y$; freezing, when $\gamma_Q=0,x<<y$; tracking,
when $\gamma_Q=\gamma_B\beta/(\beta+2)$ and $(y/x)^2\equiv
V/K=2/\gamma_Q-1\approx const$.

Besides the roll $\lambda$, the potential depends on the curvature
$\eta\equiv d^2 V/V(\varkappa d \phi)^2$.  When
the roll is flat
($\epsilon\equiv\lambda^2/2 \ll 1$), the kinetic energy
$\dot\phi^2 /2$ is negligible in the quintessence energy (12). When
the curvature is
small ($\eta\ll 1$, $\ddot\phi$ is negligible in the equation of
motion (10).
In ordinary inflation, both these conditions hold ({\em slow
roll approximation}): the expansion is dominated by the
cosmological drag and the field is nearly frozen.
But, in
quintessence, the acceleration began only recently, so that the
roll $\lambda_0$ and curvature $\eta_0$ are still $\mathcal{O}(1)$. This invalidates the
slow roll approximation for quintessence, so that the dynamical
equations need to be integrated numerically. We ultimately handled
the kination/freezing and freezing/tracking transitions and numerical
stability and round-off problems in the protracted frozen era, by implicit
Adams backward differentiation procedures (Maple lsode[adamsfunc]
and lsode[backfull]), with small adaptive step-size.

Unless it undergoes a first-order phase transition, the
quintessence field rolls monotonically towards a minimum at
$\phi=\infty$ or at some finite $\phi_{min}$: either way, the
potentials we consider are always convex. (Presumably, there is no
true cosmological constant so that the potential energy vanishes
asymptotically, avoiding the possibly worrisome future 
de Sitter event horizon, with attendant diminution of causal connectivity in
the far future.)
Defining $\Gamma\equiv V d^2 V/d \phi^2/(d
V/d\phi)^2$, so that $\eta=\lambda^2\Gamma$,
we have $ 1/\beta(\phi)\equiv\Gamma-1=\lambda^2 d^2\ln V/\varkappa^2d\phi^2 =
d(1/\lambda)/\varkappa d\phi>0$.  

Both the roll
and the curvature, $\lambda,~\beta$ are listed in Table I, for
five different potentials. The first, third and fourth rows in
Table I list three potentials with constant inverse curvature
$1/\beta$: the cosmological constant, inverse power, and
exponential, for $\beta=0,const\equiv\alpha,\infty$ respectively.
On the second row, where
$\tilde{\alpha}\equiv\sqrt{\gamma_Q}/\alpha$, the {\em constant
$\dub$ model} interpolates between the inverse power potential,
when $\tilde{\alpha}\varkappa\phi\ll 1$, and the exponential when
$\tilde{\alpha}\varkappa\phi\gg 1$.  The bottom row in Table I is
the more realistic inflaton broken-SUSY SUGRA potential, in which $\beta(\phi)$
decreases significantly for $\phi \gtrsim M_P $. 
%%%%%%%%%%%%%%%%%%%%%%%%%%%%%%%%%%%%%%%%%%%%%%%%
\subsection{Phaseportrait in Terms of Quintessence Kinetic, Potential
Energy Canonical Variables}

In place of the phase variables $\phi,~\dub\equiv (P/\rho)_Q ,$ we
may use  $x\equiv(\varkappa d\phi/d N)/\sqrt{6}~,
y\equiv\sqrt{V/\rho}$            , for which the equations of
motion are \cite{Fer,Liddle,Cope,Ng} \bea
dx/dN &=& -3x + \lambda\sqrt{3/2}y^2 + 3x\gamma/2 \\
dy/dN &=&   \qquad    - \lambda\sqrt{3/2}xy+ 3y\gamma/2 \\
d\lambda/dN &=&-\sqrt 6 \lambda^2 x/\beta \qquad \mbox{or}\qquad
d(1/\lambda)/d N=\sqrt 6x/\beta.\eea 
\input{TableI.tex}
The overall equation of state of our two-component mixture of
background and quintessence, $\gamma=\gamma_Q\omq+\gamma_B\ombe=
2x^2+\gamma_B (1-x^2-y^2), $ is a time-dependent function of the
scalar field $\phi(N)$. Thus, \be x^2+y^2=\omq,\quad 2 x^2=\omq
\gamma_Q,\qquad
 y^2/x^2=V/K=(1-\dub)/(1+\dub),\qquad
d\ln (x^2/y^2)/dN=6(\Delta -1).\ee The three-element system
(17-19)) is autonomous, except for the slow change in
$\gamma_B(N)$ from 4/3 to 1, while gradually going from the
radiation-dominated to the matter-dominated universe, around
red-shift $z_{eq}=3454$ .

The magnitude of $V$ needs to be fitted to the present value $V_0=\rho_{cr0}
y_0^2=\rho_{Q0}(1-w_{Q0})/2$. For example,
inverse power potentials, require energy scale
$M_{\alpha}=(V_0\phi_0 ^\alpha)^{1/(4+\alpha)}$, listed in the
fourth column of Table II. For shallow potentials
($\alpha<0.2$), this energy scale is close to observed neutrino
masses and to the present radiation temperature, possibly
suggesting some role for the neutrino mass mechanism or for the
matter/radiation transition, in bringing about quintessence
dominance. For steep potentials ($\alpha>1$), this mass scale can
be considerably larger, suggesting the larger scales we encounter
in particle physics.

While the evolution of a
homogeneous scalar field depends only on its equation of state
$\dub=(P/\rho c^2)_Q$, the growth of its fluctuations depends also
on the quintessence sound speed $c_s^2=(d P/d \dot\phi)/(d\rhoq/d
\dot\phi)$. With the linear form for the kinetic energy
$K=\dot\phi^2/2$ that canonical quintessence assumes, $c_s^2=c^2$ and
$-1\leq \dub\leq 1,~d \dub/dz>0$. Non-canonical forms, as in 
k-essence \cite{Armen,Erikson}, would allow $\dub<-1,~d \dub(z)/dz<0$ and give different sound speed and structure
evolution. Despite this difference in sign of $d\dub/dz$,
k-essence is hardly distinguishable from quintessence, unless
$c_s^2\approx 0$ since the surface of last scattering
\cite{Barger}. We will consider only canonical quintessence
evolution with the Friedmann expansion rate (2).
\input{QuintrackMykanos.tex}

\input{InvPowerMykanos.tex}

\input{ConclusionsMykanos.tex}

\begin{acknowledgments}
We are indebted to Martin Block for help with the numerical
integration of the scalar field dynamical equations and to US
Department of Energy grant DE-FGO2-95ER40893 at the University of
Pennsylvania for partial travel support.
\end{acknowledgments}

\input{bibliographyMykanos.tex}
\end{document}

%% file: TableI.tex
\begin{table}
\caption{Potentials described by roll $\lambda=-d\ln V/\varkappa
d\phi$ and curvature $\eta=d^2 V/V d(\varkappa \phi)^2$.}
\begin{ruledtabular}
   \begin{tabular}{|l|c|c|c||r|}
   {\bf $V(\phi)$ }                        &{\bf $\lambda(\phi)$}
   &{\bf $\eta(\phi)=\lambda^2\Gamma$} &{\bf $\Gamma-1=1/\beta(\phi)$} &NAME \\  \hline
   $\exp{-\lambda\varkappa\phi}$
   &$\lambda=const>\sqrt{3\gamma_B}
   $ &$\lambda^2=const>3\gamma_B$        &$0$  &exponential \\
   $1/\sinh^{\alpha}(\tilde{\alpha}\varkappa\phi)$
   &$(\alpha\tilde{\alpha})\coth(\tilde{\alpha}\varkappa\phi)$
   &$(\alpha\tilde{\alpha})^2 [(1+\alpha)\coth(\tilde{\alpha}\varkappa\phi)-1]$
   &$1/\alpha\cosh^2(\tilde{\alpha}\varkappa\phi) $
   &const $\dub=-2/(2+\alpha)$ \\
   $\phi^{-\alpha}$ &$\alpha/\varkappa\phi$
   &$\alpha(\alpha+1)/(\varkappa\phi)^2$     &$1/\alpha$
   &inverse power \\
   $const$ &0  &0 &$\infty$  &cosmological const \\
   $\phi^{-\alpha}\cdot\exp{\frac{1}{2}(\varkappa\phi)^2}$
   &$\alpha/\varkappa\phi-\varkappa\phi$
   &$[\alpha(\alpha+1)-\alpha(\varkappa\phi)^2+(\varkappa\phi)^4]/(\varkappa\phi)^2$
   &$(\alpha+(\varkappa\phi)^2)/(\alpha-(\varkappa\phi)^2)^2$ & SUGRA
   \end{tabular}
\end{ruledtabular}
\end{table}

%% file: QuintrackMykanos.tex
\section{ATTRACTORS IN BOTH TRACKING AND IN QUINTESSENCE-DOMINATED ERAS}

The equations of motion enjoy a fixed attractor solution
$\lambda=const$ (exponential potential), when $\beta=\infty$, and
an {\em instantaneous attractor} solution
\cite{Wett,SWZ,ZWS,Cope}, when $\lambda\beta\gg 1$ makes $\lambda$
slowly varying in equation (19) and the equation of state becomes
\cite{Ng}
\[
\gamma_{Qatt}(\lambda)=\frac{1}{2}[((1+2/\beta)t+\gamma_B)-\sqrt{((1+2/\beta)t-\gamma_B)^2+8\gamma_B
t/\beta}]=\left\{\begin{array}{ll} 0                 , &\beta=0
  (\mbox{cosmological const}) \\
                                      t-(1-2/\beta^2)t^2/2\gamma_B , & t \ll 1 (\mbox{quintessence era)}\\
                                    \gamma_B\beta/(\beta+2)        , &
                                    t \gg 1(\mbox{background era}),\\
                                    \gamma_B                    ,&\beta=\infty (\mbox{exponential potential}).
                   \end{array} \right.
\]
where $t\equiv \lambda^2/3$. 
Defining
$\Omega_{Qatt}(\lambda)=\gamma_{Qatt}/t,~
\lambda_{att}\equiv\sqrt{3\gamma_{Qatt}/\omq}$, so that
$\Delta =\lambda/\lambda_{att}=\sqrt{\gamma_Q/\gamma_{Qatt}}$, the
Attractor Condition is simply \be \Delta\approx 1 \qquad \mbox{or}\qquad
\lambda \approx \lambda_{att}.\ee
Quintessence \cite{Ratra,Wett,SWZ} exploits this attractor property to
explain
the very small present smooth energy density dynamically, without
invoking a finely tuned cosmological constant.

We reserve the term {\em tracker} for  
attractors in the background-dominated era, for which $\beta$ and $\gamma_Q=2
\gamma_B/(2+\beta)$ are nearly constant, so that $\Omega_{Qtr}\sim a^{3(\gamma_B-\gamma_Q)} \sim t^P$,
where $P=4/(2+\beta)$.
{\em Good trackers} ($\beta>1,~\gamma_Q\lesssim \gamma_B)$, have $\omq$ and
$\lambda_{att}$ slowly varying. Very good trackers ($\beta\gg 1$) approximate exponential potentials;
because $\omq (z)$ was still appreciable in the distant past, they 
show {\em early quintessence}. In {\em poor trackers}
($\beta<1,~\gamma_Q\ll 1$),  $\omq\sim t^2,
~\lambda_{att}\sim t^{-1}$ are changing with time and $\eta$ can be
appreciable, even when the roll $\lambda$ is small.

Figure
1 shows the complete attractor phaseportraits, while tracking and
while quintessence-dominated, for three forms of potential: constant
$\dub$ (dashed), inverse power (solid), SUGRA (dotted)), all
chosen to reach $\Omega_{Q0}=0.71$ at the present time. For each
potential, the shallow potential $\alpha=1$ attractor trajectories
appear on the left, the steep $\alpha=6$ attractor on the right.
The cosmological constant trajectory $\alpha=0$ is the y-axis. Not
shown in the figure, is
the exponential potential's fixed point,
$x=\sqrt{\Omega_{Q0}/2}=y$,
lying on
the circle $\Omega_{Q0}=0.71$. The presently observed trapezoidal 
region in phase space marginally allows the $\alpha=1$ inverse
power potential, but comfortably allows the SUGRA potential for a
range in parameter $\alpha$. 

In the
background dominated era, all these potentials track with phaseportrait slope $(y/x)_{tr}=\sqrt{V/K}$ gradually
increasing from $\sqrt{1/2+3/\beta}$ in the radiation-dominated era
to $\sqrt{1+4/\beta}$ in the matter-dominated era. Later, as
quintessence begins dominating, this slope gradually steepens, so that
the phase trajectories slowly approach the x-axis.  At present, we are
now
past tracking but not yet de Sitter.
For most potentials,
$\phi_0\sim\ M_P$, so that we need to consider 
trajectories whose present roll and curvature are $\mathcal{O}(1)$. 
After quintessence
begins dominating, their different curvatures make these three
potential forms evolve differently: the constant $\dub$ potential curves keep their
tracker values, the SUGRA potential orbits curve strongly towards
the y-axis (cosmological constant). 
\input{figures1.tex}

\input{TableII.tex}
\input{figures2.tex}

%% file: figures1.tex
\begin{figure}[t!]
\includegraphics*[width=14cm]{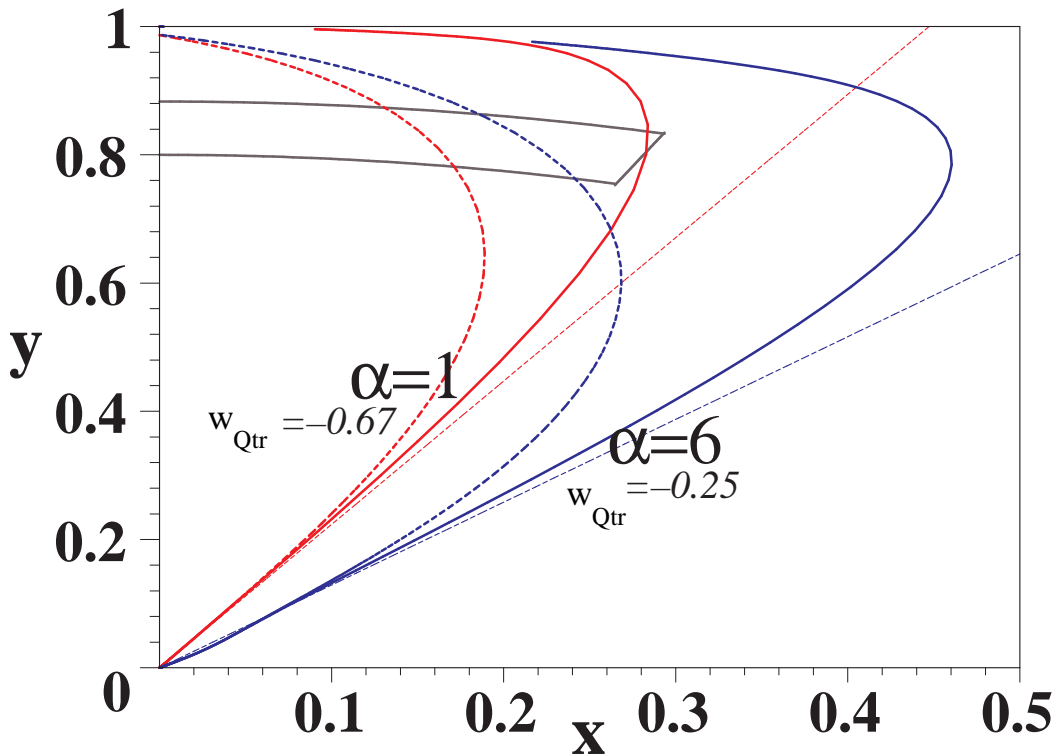}
\caption{\label{Fig1.eps} Complete phaseportraits of attractors
now passing through $\Omega_{Q0}=0.71$, for different curvature
potentials: three steep potentials($\alpha=6,w_{Qtr}=-0.25$) on the right,
three shallow potentials ($\alpha=1,w_{Qtr}=-0.67$) on the left.  In the
background-dominated era $\Omega_Q=x^2+y^2\ll 1$, all attractors
are nearly linear (track). At any  $\Omega_Q$, shallow
potentials are poorer trackers than are steep potentials. In the
quintessence-dominated era, different
curvature effects emerge: both constant $w_Q$ attractors (dashed)
remain linear; both inverse power attractors (solid), start
late to curve slowly towards the y-axis ($w_Q=-1$); both 
SUGRA attractors (dotted), start early to curve rapidly towards the
y-axis. The trapezoidal region on the upper left is the
observationally allowed present phase space.}
\end{figure}
%%%%%%%%%%%%%%%%%%%%%%%%%%%%%%%%%%%%%%%%%%%%%%%%%%%%%%%%%%%%

%% file: TableII.tex
\begin{table}
 \caption{Tracker and present ($\Omega_{Q0}=0.71$) attractor solutions for inverse power potentials.}
\begin{ruledtabular}
   \begin{tabular}{|r||c|c|c||c|c|c|c|c|c||c}
$\alpha$ &$w_{Qtr}$      &log $\Omega_{Qtri}$ &$M$ &$w_{Q0}$
&$x_0^2$ &$y_0^2$ &$\eta_0$       &$\lambda_0$ &$\varkappa\phi_0$
&log $\Omega_{Qi}$ range
\\  \hline
   6     &   0..-0.25      &-10.9   & 5.3~PeV    &-0.41  &.210   &.499
   &2.69 &1.519(fast)       &3.949    &-42..-0.5 \\
   1     &  -0.555..-0.667 &-29.3   & 2.3~keV        &-0.76    &.083   &.626
   &1.65 &0.908       &1.101    &-41..-1.5\\
   0.5   &  -0.733..-0.80  &-35.2   & 4.8~eV     &-0.86  &.049   &.661
   &1.42 &0.689             &0.73     &-42.1..-9.1\\
   0.1   &  -0.937..-0.952 &-41.9   &12.1~meV    &-0.97  &.011   &.698
   &1.36 &0.351(slow )&0.285    &-42.2..-35.5\\
   0     &  -1             &-44.1   & 2.5~meV    &-1     &0      &.71
   &0    &0 (static)        & -       &-44.1
   \end{tabular}
\end{ruledtabular}
\end{table}
%%Include w_1\equiv (d\dub/dn)_0=3(1-\dub0^2)(\Delta-1)/

%% file: figures2.tex
%\documentclass[aps,prd]{revtex4} %figures2 Changing Windows landscape->portrait
%to avoid rotation doesn't change BB, which can be read off .eps-i.
%Original *.eps file is not changed.Therefore rotation will still be needed.
%BB is ultimately chosen in \includegraphics.
%\usepackage[dvips]{graphicx}
%\usepackage{amssymb}
%\begin{document}

\begin{figure}[t!]
\includegraphics*[bb=72 72 540 719,angle=-90,width=8.5cm]{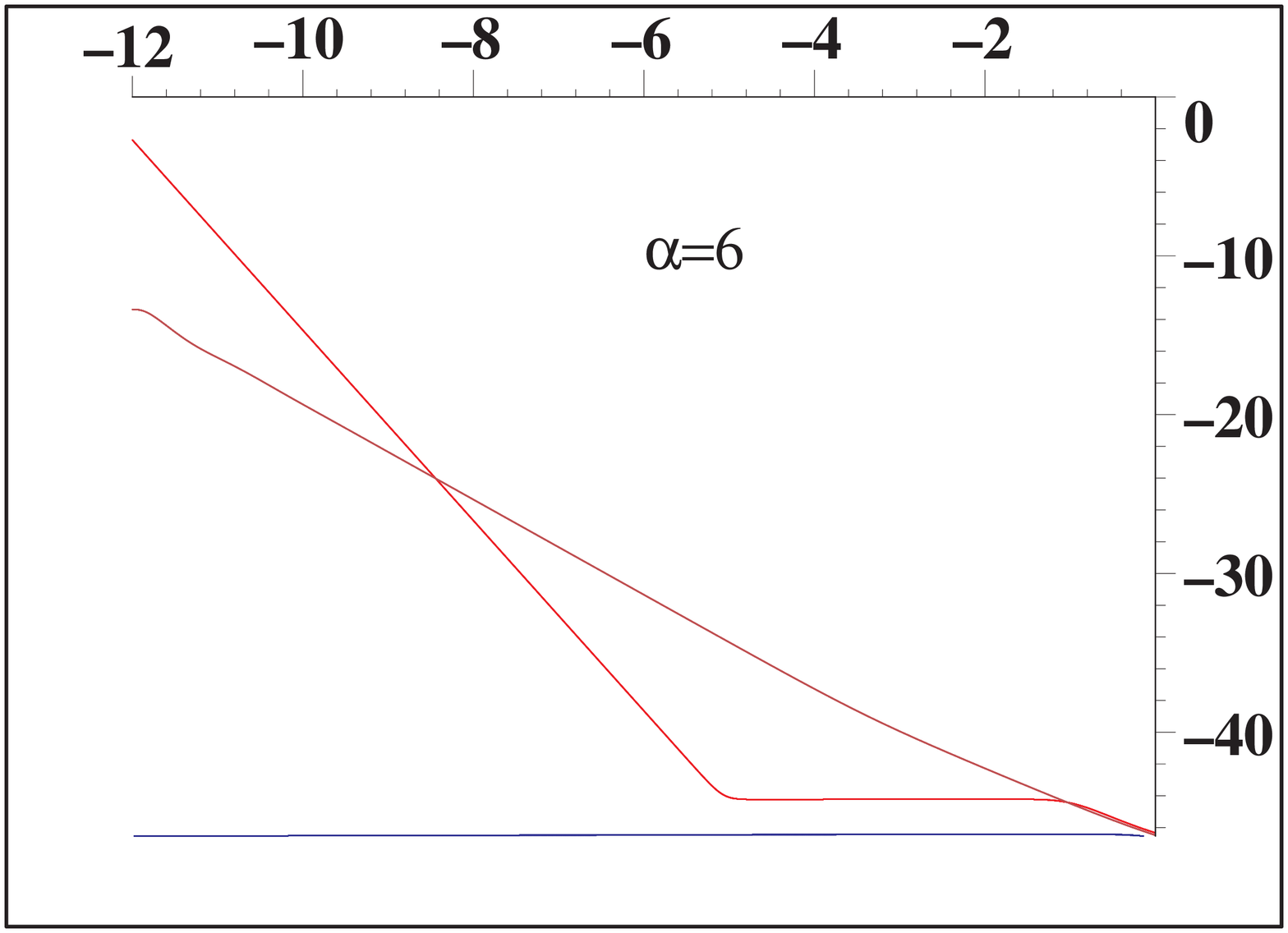}
\includegraphics*[bb=72 72 540 719,angle=-90,width=8.5cm]{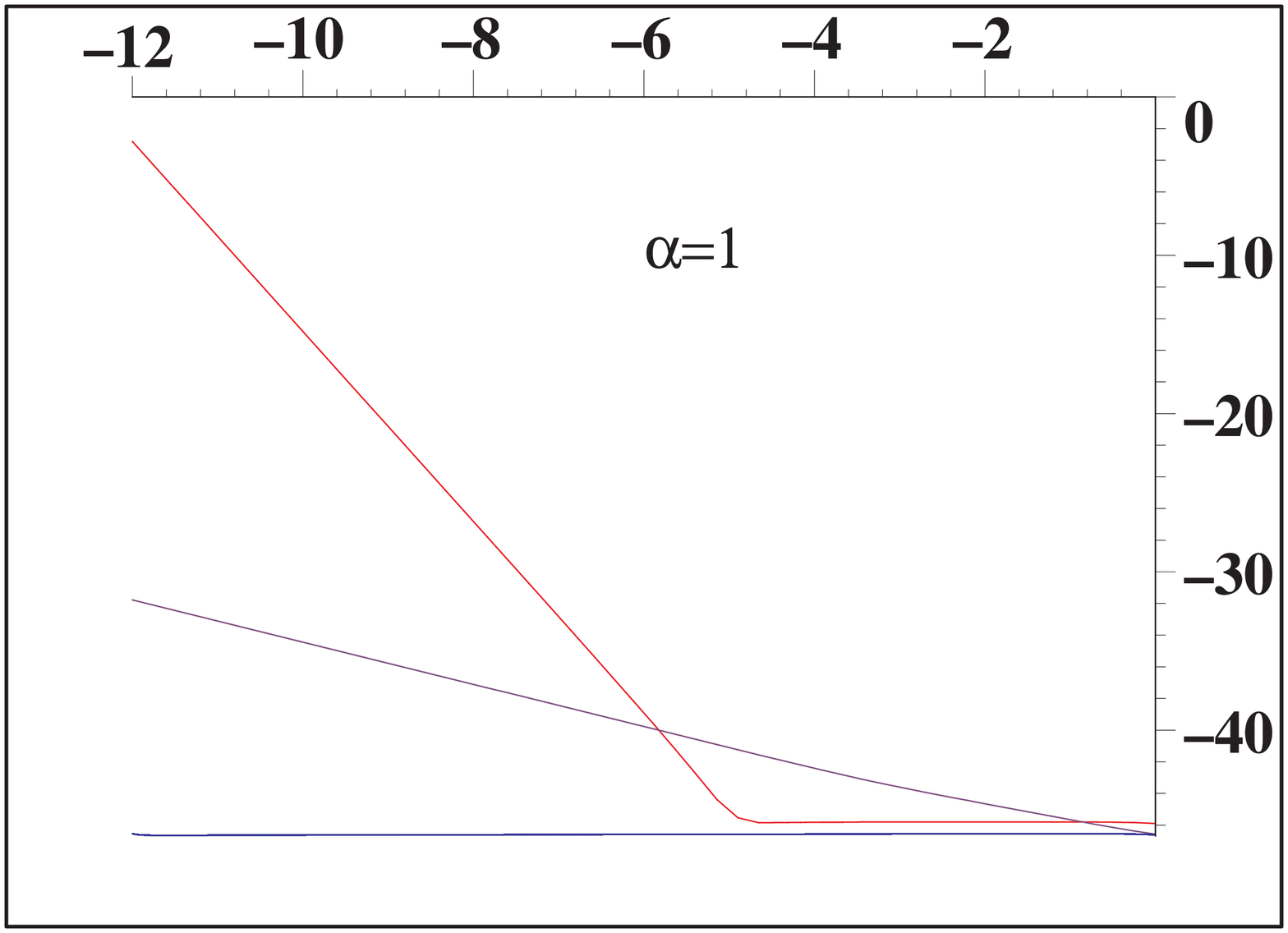}
\includegraphics*[bb=72 72 540 719,angle=-90,width=8.5cm]{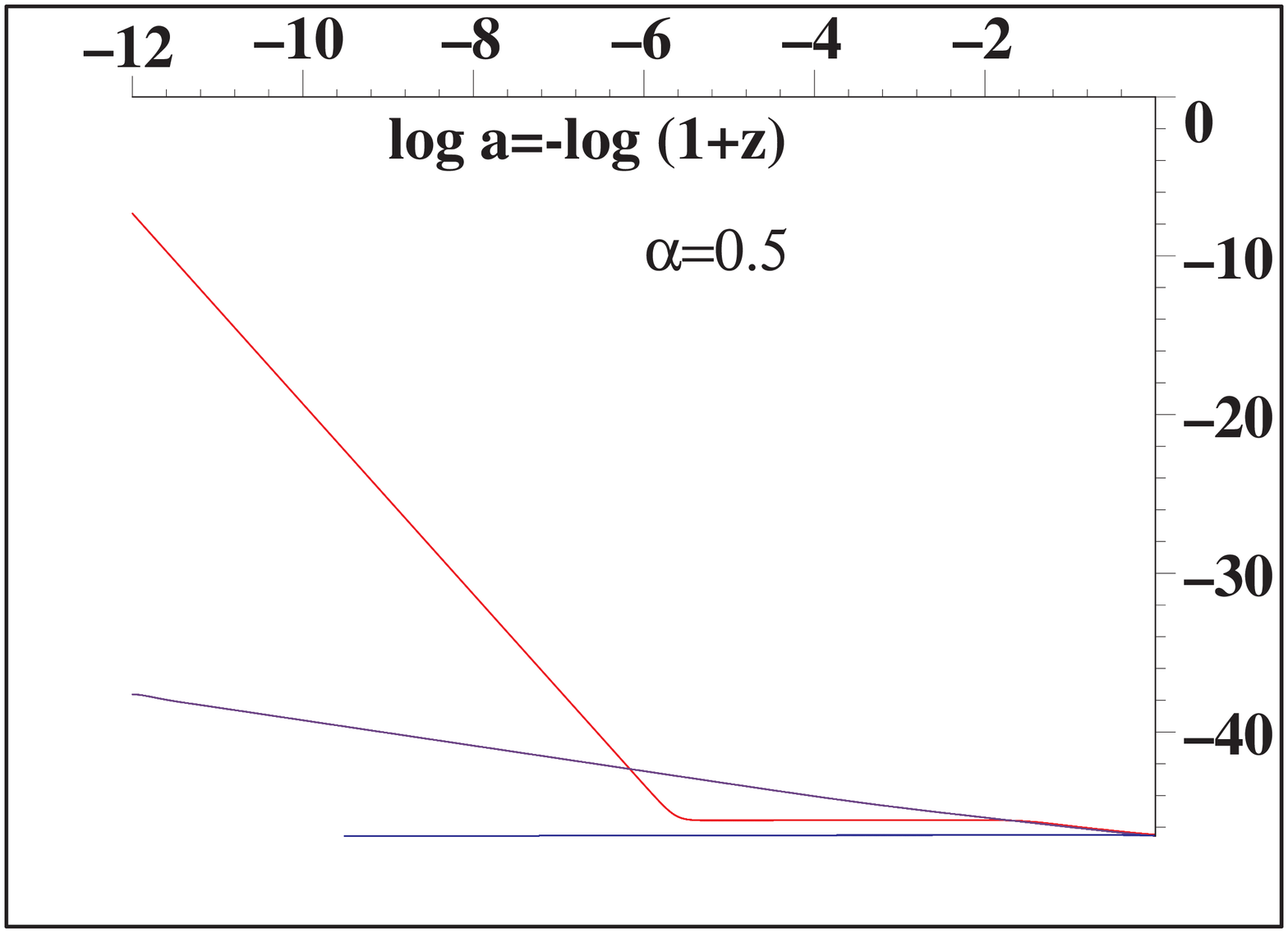}
\includegraphics*[bb=72 72 540 719,angle=-90,width=8.5cm]{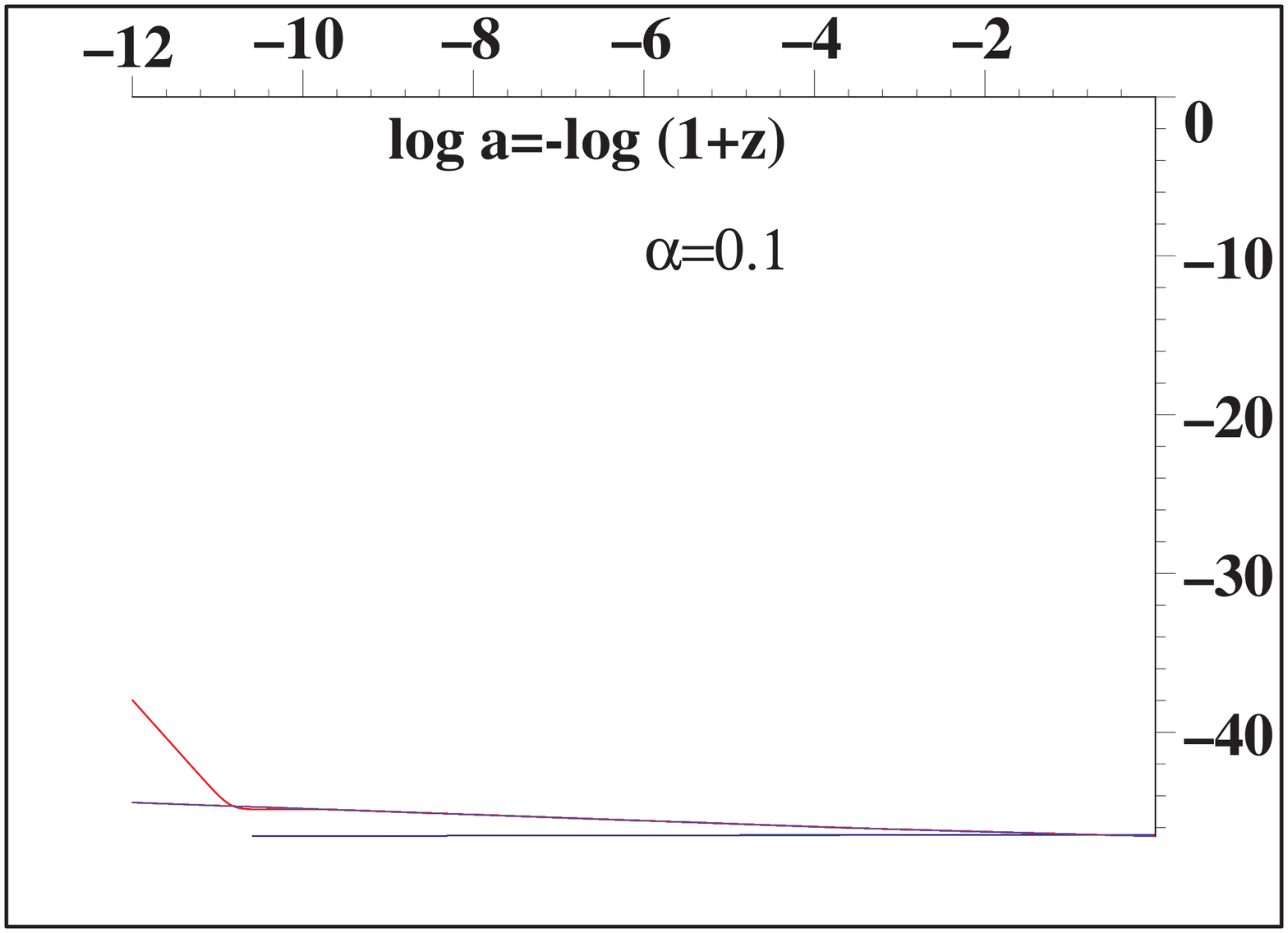}

 \caption{\label{Fig2.eps}
Evolution of quintessence energy density, log $\rho_Q (GeV^4)$ on
vertical axes, for
four inverse power potentials $\alpha=6,1,0.5,0.1$, from red-shift
$z=10^{12}$, to the present value $\rho_{Q0}=0.71\rho_{cr0}$. In
all figures, the central, trajectory is the attractor, starting
with tracker slope $d \ln\rho_Q/d N=-6\gamma_B/(2+\alpha)$. The
lower curve is the maximal undershoot trajectory, which freezes
immediately and then crawls slowly to join the attractor now. The
upper curve is
the maximal overshoot trajectory, which kinates with slope -6, before
freezing late and now reaching the attractor.  Poor trackers must
freeze early, out of a narrow range in log $\rho_Qi$. }

\end{figure}

%%%%%%%%%%%%%%%%%%%%%%%%%%%%%%

%\end{document}

%% file: InvPowerMykanos.tex
\section{SHALLOW INVERSE POWER POTENTIALS MAKE POOR TRACKERS}

To obtain the presently small smooth energy without fine-tuning
initial conditions, phase trajectories must flow onto the
attractor before now and for a broad range of initial conditions,
the {\em basin of attraction}. A good tracker starts from a broad
basin of attraction and freezes with $\dub\sim -1$,
before tracking with kinetic energy/potential energy $\sim 1$.  A poor
tracker starts from a narrow
basin of attraction and freezes for a long time, before tracking with
kinetic energy/potential energy $\ll 1$. 

In poor attractors, this shrinkage of the basin of attraction, making
the present universe sensitive
to initial conditions, derives
from early freezing. From any {\em undershoot} initial 
conditions
$\rho_{Q0}<\rho_{Qi}<\rho_{Qtr}$, freezing emerges
directly. (Indeed, if the initial density
$\rho_{Qi}$ is small enough , the field crawls down to its present
value $\rho_{Q0}$, without ever tracking.) But, starting from {\em
overshoot} initial conditions $1>\Omega_{Qi}>\Omega_{Qtr}$,
freezing starts much later, at a value $\phi_{fr}\gtrsim M_P
\sqrt{6\Omega_{Qi}}$, only after a long kinated era ($x \gg y$), during
which $\dot{\phi}^2\approx 2\rhoq$ lets the field grow rapidly
while $\omq$ decreases. Too much overshoot would make the phase
trajectories freeze so late as to reach the attractor only in the
future. 

We must now extend earlier treatments \cite{ZWS,SWZ} of trackers
to solutions that reach the
attractor only now, in the present, quintessence-dominated era.  When the curvature is large,
tracking will stop early, while $\lambda\sim\beta$ is
already slow-rolling. Attractors that only now become approximately slow-rolling
($\lambda_0 < 1$) could only have been reached 
from a narrow tracking basin of attraction. Indeed, in the static
limit (cosmological constant), the present smooth energy arises
out of the unique initial condition $\rho_{Qi}=\rho_{Q0}$. 

For the simple inverse power law potentials \be
V(\phi)=M_{\alpha}^{4+\alpha}/\phi^\alpha,\ee the curvature
$\beta=\alpha= constant$, so that the equation of motion (10) has
exactly scaling solutions in both the radiation- and the
matter-dominated eras. These potentials are interesting,
because they approximate any potential, while tracking. They arise naturally in supersymmetric
condensate models for QCD or instanton SUSY-breaking
\cite{Bin,Brax,Brax2}, but aquire appreciable quantum corrections
when $\phi\gtrsim M_P$. For these potentials, $\lambda\sim
V^{1/\alpha}\sim (yH)^{2/\alpha}$, the third equation (19)
integrates to $\lambda=\lambda_0(yH/y_0 H_0)^{2/\alpha}$ in terms
of present values of $y,~H,~\lambda$.

The second column in Table II gives the range in tracker values
$w_{Qtr}$, from $(\alpha-6)/3(\alpha+2)$, during the radiation-dominated era, to
$-2/(\alpha+2)$, during the matter-dominated era.  The third
column gives the initial $\Omega_{Qtri}$ values at $z=10^{12}$ a
tracker must have in order to reach the present value
$\Omega_{Q0}=0.71$. The fourth column tabulates the quintessence
energy scales needed, in order to reach $\Omega_{Q0}=0.71$. 

After tracking, these
trajectories curve towards the $x=0$ (asymptotic de Sitter) axis. Between the two vertical double bars, columns five  through ten,
summarize the present (post-tracker) values for trajectories that
have tracked before now. The steep $\alpha=6$ trajectory
would by now track down only to $w_{Q0}=-0.41$, which is excluded
observationally by the constraint (8), which requires
$\alpha\lesssim 1$ \cite{SWZ}. 

Integrating equations (17,18) shows, in the
last column, for each $\alpha$, the range of initial $\Omega_{Qi}$
that would track and finally reach the presently-allowed value
$\Omega_{Q0}=0.71$.
For large $\alpha$, the quintessence potential would now still be
fast-rolling \cite{Blud}. As $\alpha$ decreases, $w_{Q0}$
decreases, but the basin of attraction shrinks. For $\alpha<0.5$,
the presently-tracking range in initial values of log $\Omega_{Qi}$ 
is already 9 orders of magnitude narrower than the
initial range of the good $\alpha=6$ trackers first considered
\cite{SWZ}. For a cosmological constant ($\alpha=0$), the present
value is realized only if it is initially tuned uniquely to its
present value $\rho_Q=28.8~meV^4$.

Because the observed  $\widetilde{\dub}$ is already close to the
cosmological constant value $-1$, an inverse power potential requires $\alpha<1$, so
that the potential energy {\em always} dominated the kinetic
energy ($x<<y$).
These nearly flat trajectories never track, but "crawl"
\cite{Huey} towards their present values, only because they were
initially tuned close to these values. For a good tracker to
nevertheless reach
$\widetilde{\dub}<-0.78$, its post-tracker potential needs $\beta(\phi)$ decreasing, $P(\phi)$ increasing, so
that $\omq\sim t^P$ grows rapidly at late times \cite{SWZ}.
Such {\em cross-over quintessence} \cite{Cald} is characterized by $\dub (z)$
reduction in the recent past ($z<0.5$).
\input{TableIII.tex}

%% file: TableIII.tex
\begin{table}
\caption{Quintessence potentials which track early, but crawl
now.}
\begin{ruledtabular}
   \begin{tabular}{|l|c|r|}
   {\bf Potential $V(\phi)$}                                       &{\bf Theoretical Origin}                           &{\bf References}\\  \hline
    $M^4[\cos(\phi/f)+1]$                                          &String, M-theory pseudo Nambu-Goldstone light axion&\cite{Frieman2,CDF,CDS,Choi}\\
    $M^{4+\alpha} \phi^{-\alpha} \cdot\exp{\frac{1}{2}(\varkappa\phi)^\beta/2}$&SUGRA, minimum at $(\varkappa\phi)^\beta=2\alpha/\beta$                                       &\cite{Bin,Brax,Brax2}      \\
    $M_P^4 [A+(\varkappa\phi-\varkappa\phi_m)^{\alpha}]\exp(-\lambda\varkappa\phi)$              &Exponential modified by prefactor, to give local minimum; M-theory &\cite{Albr,Skor}
   \end{tabular}
\end{ruledtabular}
%% Omit \cite{Huey2}?; Are Weller's references cited for Strings correct?
\end{table}

%% file: ConclusionsMykanos.tex
\section{ONCE GOOD TRACKERS NOW ROLLOVER TOWARDS LARGE POTENTIAL CURVATURE}

We have not considered models with a true cosmological constant,
matter-coupled quintessence, quantum corrections to classical quintessence,
k-essence \cite{Armen,Barger}, nor large extra
dimension or brane models \cite{Nunes,AlbBurg}, for
which the Friedmann equation is modified at very early times.
But, within canonical quintessence,  
the observations allow
phase trajectories that are insensitive to initial
conditions, only if the curvature increases rapidly near
the present epoch $\varkappa\phi_0\sim 1$, so that
quintessence-domination and kinetic energy suppression began only recently.

This needed post-tracking behavior is illustrated in the popular
potentials listed in Table III.  (A longer list of potentials is
given in references \cite{Well,Sahni,Peebles,Starob}.) For example, the SUGRA
models on the last line of Table I have minima at
$\varkappa\phi=\sqrt\alpha$. The dotted curves in Figure 1 show that
far below this minimum, they behave
and track like inverse power potentials, but near the minimum, the
curvature increases rapidly. After
tracking in the background-dominated era, these SUGRA phase
trajectories,  for
$\alpha=6$ and $1$, both curve over to lower $w_{Q0}$ values, in marginal
agreement with observations, for a large range in $\alpha$ values.

A
good tracker, insensitive to initial conditions, requires a rapidly changing
potential curvature, making $\widetilde \dub (z)$ fast-changing
only recently, at red-shifts $z \lesssim 0.5$. Difficult combined
supernova, CBR and cosmic shear observations in the next decades
may yet distinguish quintessence from a true cosmological constant 
\cite{Well,Frieman}. Otherwise, quintessence appears
hardly distinguishable, theoretically and phenomenologically, from
the small cosmological constant it was designed to explain.

The original cosmological coincidence problem was to understand why
the smooth energy density is {\em now} so small, fortunately
allowing large scale structure formation, before lately dominating
the matter energy density. By requiring that the smooth
energy density now be small {\em and} fast-changing, this coincidence 
problem is now exacerbated. In two ways. 
recent cosmological observations stress the special time in which
we live, which may call for anthropic reasoning in cosmology.